\documentclass[]{spie}  

\def\gtrsim{\mathrel{\hbox{\rlap{\hbox{\lower4pt\hbox{$\sim$}}}\hbox{$>$}}}}
\def\lesssim{\mathrel{\hbox{\rlap{\hbox{\lower4pt\hbox{$\sim$}}}\hbox{$<$}}}}

\usepackage[]{graphicx}

\title{A Concept for a High-Energy Gamma-ray Polarimeter} 

\author{P. F. Bloser\supit{a}, S. D. Hunter\supit{a}, 
G. O. Depaola\supit{b}, and F. Longo\supit{c}
\skiplinehalf
\supit{a}NASA Goddard Space Flight Center, Code 661, Greenbelt, MD 20771, USA 
\skiplinehalf
\supit{b}Facultad de Matem\'atica, Astronom\'\i a y F\'\i sica, Universidad Nacional de C\'ordoba, 
Ciudad Universitaria, 5000 C\'ordoba, Argentina 
\skiplinehalf
\supit{c}Department of Physics and INFN, via Valerio 2, I-34100 Trieste, Italy
}

\authorinfo{Further author information: (Send correspondence to P. F. Bloser)\\P.F.B.: E-mail: bloser@milkyway.gsfc.nasa.gov}

\begin{document} 
  \maketitle 

\begin{abstract}

We present a concept for an imaging gamma-ray polarimeter operating
from $\sim 50$ MeV to $\sim 1$ GeV.  Such an instrument would be valuable for the
study of high-energy pulsars, active galactic nuclei, supernova
remnants,  and gamma-ray bursts.  The concept makes use of pixelized gas
micro-well detectors, under development at Goddard Space Flight
Center, to record the electron-positron tracks from pair-production
events in a large gas volume.  Pixelized micro-well detectors have
the potential to form large-volume 3-D track imagers with $\sim 100$ $\mu$m
(rms) position resolution at moderate cost.  The combination of high
spatial resolution and a continuous low-density gas medium permits
many thousands of measurements per radiation length, allowing the
particle tracks to be imaged accurately before multiple scattering
masks their original directions.  The polarization of the incoming
radiation may then be determined from the azimuthal distribution of
the electron-positron pairs.  
We have performed Geant4 simulations of these
processes to estimate the polarization sensitivity of a simple telescope
geometry at 100 MeV.

\end{abstract}



\section{INTRODUCTION}
\label{sect:intro}  

Long recognized as a potentially valuable tool for high-energy astronomy, gamma-ray
polarimetry is finally showing signs of becoming a viable observational technique. 
Traditional gamma-ray telescopes have thus far relied exclusively on imaging, 
spectroscopy, and timing to determine what physical processes are
at work in high-energy sources.  Due to low statistics, however, 
it is usually quite difficult to distinguish between competing physical models based
on spectral and timing information alone.  The promise of gamma-ray polarimetery
stems from the fact that nearly all high-energy emission mechanisms can give rise
to linearly polarized emission, though the polarization angle and degree of
polarization are highly dependent on the source physics and geometry\cite{lei}.  
Both synchrotron radiation, produced by relativisitic electrons
spiraling around magnetic field lines, and curvature radiation, caused by electrons
following field lines that are themselves tightly curved, produce linearly
polarized radiation in which the angle traces the field direction and the 
degree is independent of energy.  Compton
scattering of photons off electrons, on the other hand, produces scattered 
radiation whose polarization degree depends on energy and scatter angle.  
These processes are expected to dominate the high-energy radiation of gamma-ray
pulsars, gamma-ray bursts (GRBs), supernova remnants (SNRs), active galactic 
nuclei (AGN), and X-ray binaries (XRB).  An especially exciting possibility is the radiation 
produced by magnetic photon splitting, which may play a role in pulsars and soft gamma-ray
repeaters with particularly strong magnetic fields.  The two photons produced
should have strong polarizations in which the angles depend on energy\cite{baring}.
Thus gamma-ray polarization studies promise to provide a powerful new method for 
studying the physics of astronomical sources, just as polarization studies
at optical and radio wavelengths have been doing for decades.  

Despite this powerful scientific incentive, progress in the field of gamma-ray polarimetry
has until recently been very slow.  In principle, all three X-ray and gamma-ray interaction
mechanisms -- photoelectric absorption, Compton scattering, and pair production -- are 
sensitive to the polarization vector of the incident photon: the secondary products of
the initial interaction -- the photoelectron, Compton-scattered photon and recoil electron, 
or electron-positron pair, respectively -- are produced with momenta described by a 
$\cos^2{\theta}$ angular distribution 
relative to the incident electric field (whether the maximum lies parallel or 
perpendicular to the field depends on the process).  High-energy polarimeters, therefore,
depend on measuring the azimuthal distribution of the secondary particles.  Until
recent years, technological constraints have prevented the accurate measurement of
this distribution, severely limiting the polarization sensitivity of gamma-ray 
telescopes.  Chief amongst these constraints were the insufficient position resolution of
the detectors, which restricts the accuracy to which the azimuthal angles can be 
determined, the small number of readout channels possible with practical electronics designs,
which limits how finely a position-sensing detector may be segmented, 
and the effects of multiple Coulomb scattering of the secondary charged particles 
in detector materials, which quickly distorts their initial momentum vectors.  Due to these
factors, the two largest gamma-ray telescopes ever flown in space, COMTPEL and EGRET 
onboard the Compton Gamma-Ray Observatory (CGRO), had negligible polarization 
sensitivity\cite{lei,lei2,mattox91}.

This situation is finally changing for polarimeters based on Compton scattering and
photoelectric absorption.  In the case of Compton telescopes, major advances in
detector position resolution, event reconstruction algorithms, and
high-density readout electronics have allowed new designs that
are compact, permitting a large effective area, while maintaining good angular 
resolution and background rejection.  The IBIS instrument on INTEGRAL will have some
polarization sensivitiy for this reason\cite{stephen}, and all currently discussed Advanced
Compton Telescope (ACT) designs will naturally be highly sensitive to polarized radiation between
several 100 keV and a few MeV\cite{tigre,mega,lxegrit,nct,act}.  
Laboratory prototypes of dedicated Compton polarimeters
have already demonstrated good sensitivity\cite{grape,gipsi}.  
Technological advances have also helped the cause of X-ray polarimetry using photoelectric
absorption ($\sim$ 0.5--10 keV) 
through the emergence of micropattern gas detectors.  These detectors combine a low-density 
gas medium with pixel readouts of very fine ($\sim 100$ $\mu$m) pitch, so that the 
initial track of the photoelectron may be accurately measured before it is distorted
by Coulomb scattering.  Good sensitivity has been demonstrated in the laboratory for
X-ray polarimeters using this principle\cite{costa,black}.  Thus polarimetry at X-ray
and medium energy gamma-ray energies seems nearly ready to become reality.
The recent detection via the Compton scattering technique of very
strong ($\sim 80$\%) linear polarization from the prompt emission of a GRB (0.15--2 MeV) by the 
RHESSI satellite\cite{coburn} has moved gamma-ray polarization to the forefront of high-energy
astrophysical investigations and provided a major spur to the development of future missions.

The flurry of activity in gamma-ray polarimetry has thus far not extended to the pair production
regime ($\gtrsim 50$ MeV).  This is despite the fact that it is in practice far easier to construct
sensitive pair telescopes than Compton telescopes due to better angular resolution, simpler
event reconstruction, and lower background.  The same technological advances that have
enabled the development of photoelectric X-ray polarimeters, namely very fine spatial
resolution in a gas detector, should in principle minimize the Coulomb scattering of
the electron-positron pair and thus finally also allow the construction of
a pair telescope with polarization sensitivity.  In this paper we describe a concept for
a high-energy gamma-ray polarimeter based on gas micro-well detectors, amorphous silicon
thin film transistor arrays, and high-density interconnect and readout electronics
technology.  This design is a natural extension of a next-generation high-energy
gamma-ray telescope concept introduced previously\cite{hunter}.

\section{HIGH-ENERGY GAMMA-RAY POLARIMETRY}
\label{sect:pol}

\subsection{Basic Principles}
\label{sect:principles}

The idea of measuring high-energy ($> 50$ MeV) gamma-ray polarization by detecting the
asymmetric azimuthal distribution of the plane defined by the electron-positron pair has been 
discussed since the 1950s\cite{wick}.  Much theoretical work has been done since then
to calculate the polarization-dependent cross section and estimate the detrimental
effects of multiple Coulomb scattering\cite{maximon,kelner,kozlenkov,kotov,depaola99}.
For the case where the incident photon, electron, and positron are nearly coplanar (i.e., 
the nuclear recoil is small), the azimuthal dependence of the cross section may be
written\cite{maximon}
\begin{equation}
\label{eq:az}
\sigma(\psi) = \frac{\sigma_0}{2 \pi}[1 + P R \ \cos(2 (\psi - \psi_0))],
\end{equation}
where $\psi - \psi_0$ is the angle between the pair plane and the gamma ray's electric field 
vector, $\sigma_0$ is the total cross section, $P$ is the fractional polarization, and $R \sim 0.1$
is a numerical factor expressing the inherent asymmetry of the process.  (More recent numerical
calculations extend the cross section to non-coplanar pairs and suggest a more complex
parameterization\cite{depaola99}; see Section~\ref{sect:geant}.)  In principle, then, one may
search for polarized emission from a source by making a histogram of the recorded pair azimuth
angles and fitting a function of the form $A \ \cos(2 (\psi - \psi_0)) + B$.  The strength of 
the azimuthal modulation may be expressed in terms of the modulation factor $Q$, usually
defined as\cite{lei,novick}
\begin{equation}
\label{eq:q}
Q = \frac{N_{max} - N_{min}}{N_{max} + N_{min}} = \frac{A}{B},
\end{equation}
where $N_{max}$ and $N_{min}$ are the number of counts recorded at the maximum and minimum of
the distribution, respectively.  In practice, rather than try to calculate $R$ in Equation~\ref{eq:az},
one would determine the modulation $Q_{100}$ resulting from 100\% 
polarized radiation, either experimentally or by means of a Monte Carlo simulation.  The degree of 
polarization measured from another source is then given simply by $P = Q_{measured}/Q_{100}$.  More 
sophisticated statistical tests for the presence of azimuthal modulation have been 
developed as well\cite{mattox90}; 
the above basic formalism, however, yields a useful simple expression for the minimum detectable
polarization (MDP) for a given instrument\cite{novick}:
\begin{equation}
\label{eq:mdp}
MDP = \frac{n_{\sigma}}{Q_{100} S}\sqrt{\frac{2 (S + B)}{T}}.
\end{equation}
Here $n_{\sigma}$ is the significance level, $S$ is the source count rate in the detector, $B$ is
the background count rate, and $T$ is the observation time.  The polarization sensitivity can
therefore be improved by increasing the effective area, decreasing the background, or increasing
$Q_{100}$.  

The effects of Coulomb scattering may be approximated by replacing the factor $R$ in 
Equation~\ref{eq:az} with 
\begin{equation}
\label{eq:scat}
R^{\prime} = R \ e^{-2 \Phi^2},
\end{equation}
where $\Phi \sim 14 L^{1/2}$ is the rms change in $\psi$ (assuming a Gaussian distribution) 
due to multiple scattering
after passing through $L$ radiation lengths (RL) of material\cite{kozlenkov,mattox90,mattox91}.  
(Monte Carlo simulations have found this approximation for $\Phi$ to be overly 
pessimistic\cite{mattox90,mattox91}, but it is still useful for comparing different
telescope designs.)
Thus the scattering has the effect of exponentially lowering $Q_{100}$, or exponentially
worsening the MDP.  This is the primary reason why all pair
telescopes to date, as well as those currently planned, have negligible polarization 
sensitivity.  For example, the EGRET telescope allowed
gamma rays to convert into pairs in tantalum conversion foils $L = 0.022$ RL
thick.  Although this gave the instrument a high efficiency, the resulting rms change in
the pair azimuth in each foil was $\Phi = 2.1$ radians.  Thus $Q_{100}$ was reduced by
a factor of $10^{-4}$, making EGRET insensitive to polarization from even the 
strongest cosmic sources\cite{mattox90,mattox91}.  Currently planned instruments such as
AGILE and GLAST use the same detection method of thick conversion foils, meaning that they
too will not have a useful sensitivity to polarization.

\subsection{Advantages of a Gas Detector}
\label{sect:gas}

The design of a polarimeter with, for example, sensitivity $10^3$ times that of EGRET, i.e.
with $R^{\prime}/R = 10^{-1}$, requires, from Equation~\ref{eq:scat}, 
that the direction of the electron and positron be
determined after traversing a distance of $\sim 6 \times 10^{-3}$ RL.  This distance, about
one-fourth the thickness of one EGRET conversion foil, implies a track imaging
detector with low density in terms of RL per sample measurement distance.  A design for a
next-generation high-energy gamma-ray telescope capable of arcminute single
photon angular resolution requires a similar low-density track
imager\cite{hunter}.  It was found that arcminute resolution is
achievable only above $\sim 3$ GeV and requires a track imaging detector with density
less than $\sim 2 \times 10^{-5}$ RL per sample. Using this design, the $6 \times 10^{-3}$ RL 
would correspond to about 300 samples along
the electron and positron tracks, more than sufficient to accurately 
determine the pair momenta.  Such a low ratio of RL to sampling length can be achieved
using gas detectors with very fine spatial resolution.

%
%
%

\section{ADVANCED PAIR TELESCOPE AND HIGH-ENERGY POLARIMETER DESIGN}
\label{sect:instrument}

We have developed a concept for a high-energy gamma-ray telescope using gas micro-well
detectors to achieve good polarization sensitivity.  In this section we describe the
detectors and readout electronics and present our concept for an advanced pair 
telescope (APT).  The expected polarimetry performance is discussed in Section~\ref{sect:pred}.

\subsection{Pixelized Gas Micro-Well Detectors}
\label{sect:micro}

\begin{figure}
\begin{center}
\begin{tabular}{lr}
\includegraphics[height=7cm]{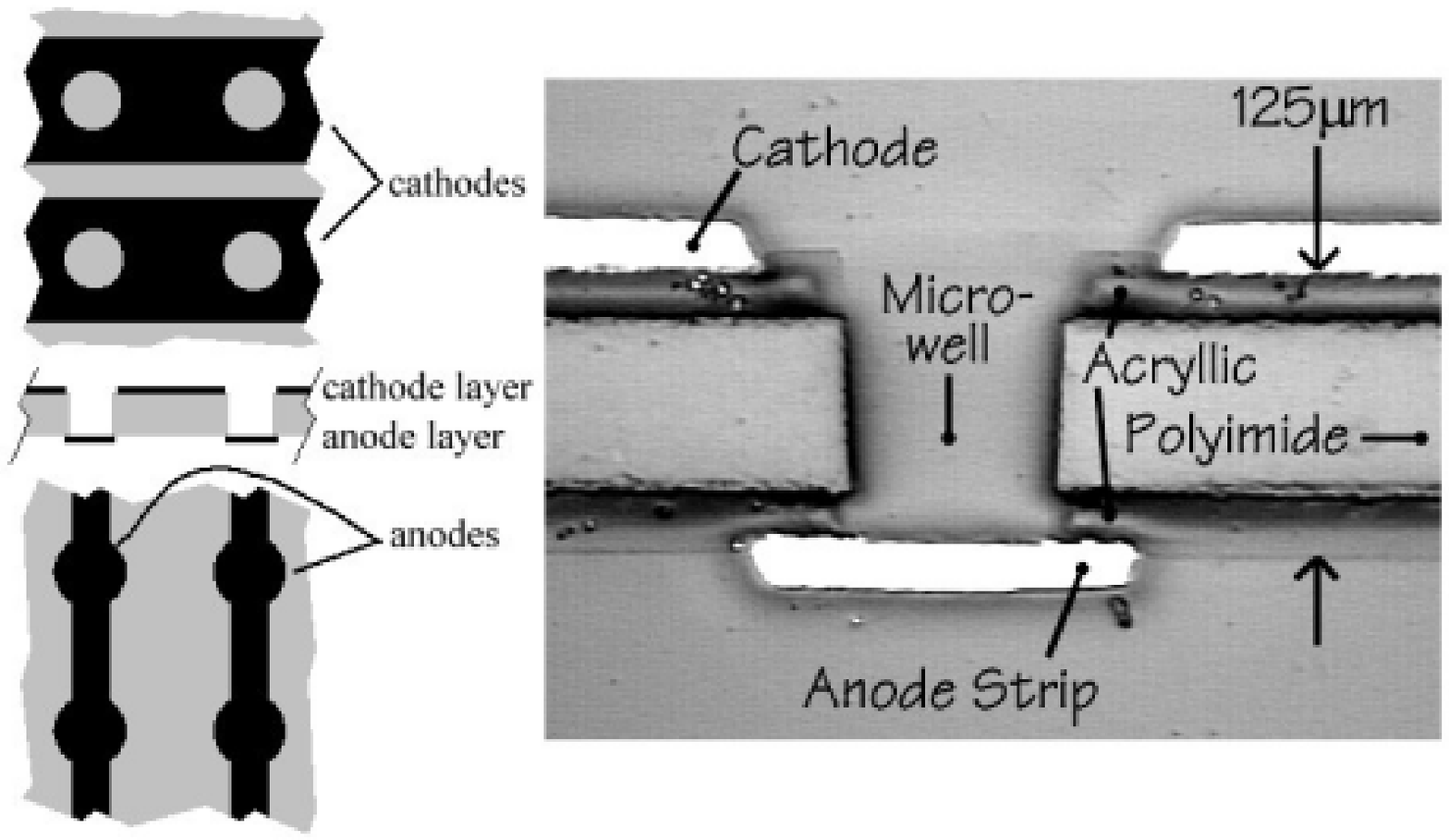}
& \includegraphics[height=7cm]{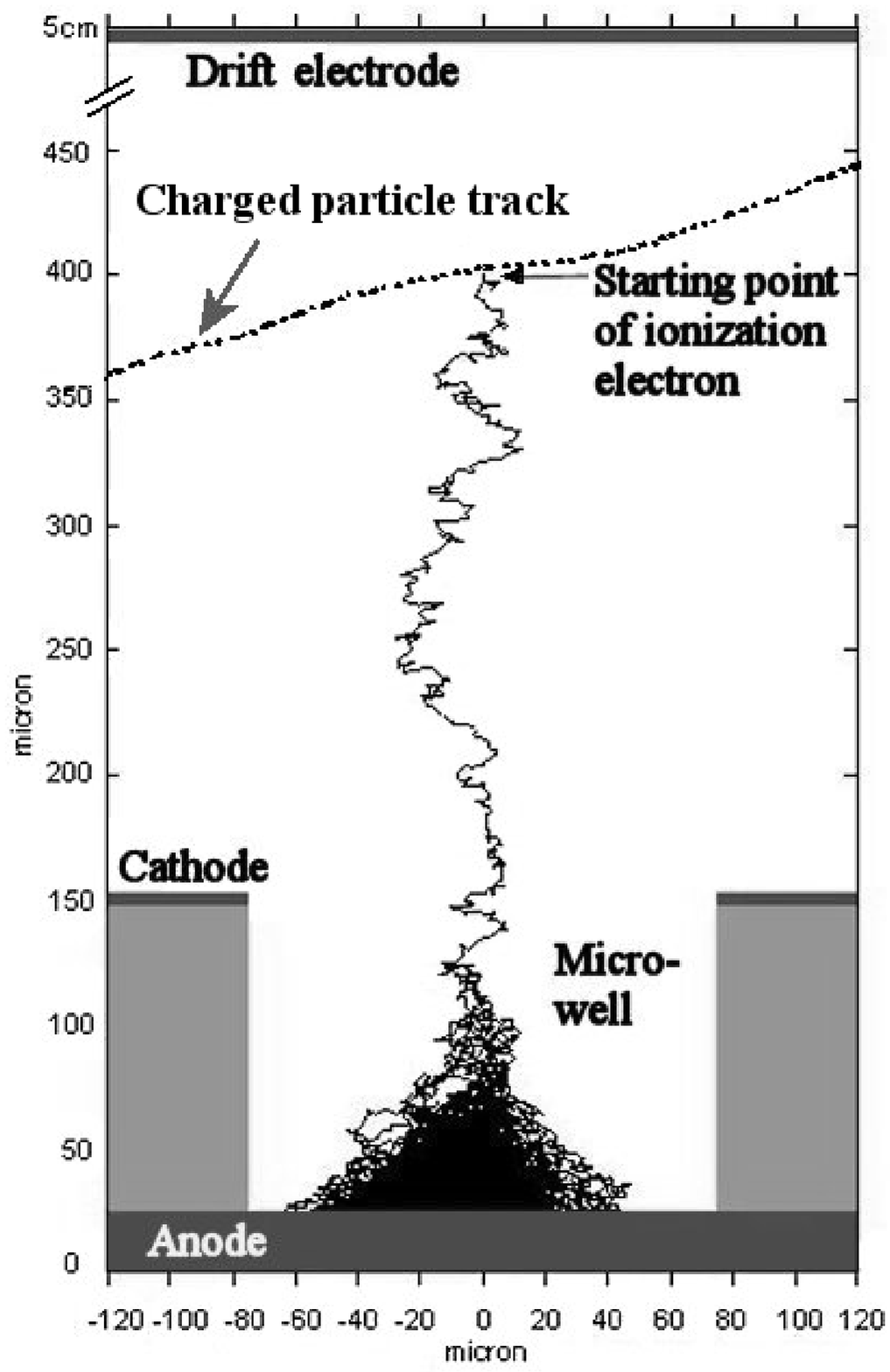}
\end{tabular}
\end{center}
\caption[example] 
{ \label{fig:well} 
Micro-well detectors.  {\em Left:} Copper cathode (top) and anode (bottom) traces for crossed-strip
design.  Wells are formed by exposing the anode through the cathode and substrate.  {\em Center:} Cross
section of an individual well formed by {UV} laser ablation with a polymide substrate\cite{black00,jones1}.  
{\em Right:} Schematic of complete micro-well detector.  The drift electrode defines the active region,
in which fast charged particles generate ionization electrons which drift into the wells.  Charge
amplification takes place in the high-field avalanche region inside each well.  The vertical scale
has been greatly compressed.}
\end{figure}

The micro-well detector\cite{black00,jones1,jones2} is a type of gas proportional counter 
related to the gas electron multiplier\cite{sauli}.  Each sensing element consists
of a charge-amplifying well, as shown in Figure~\ref{fig:well}.
The cathode and anode electrodes are deposited on opposite sides of an insulating substrate.
The well is formed as a cylindrical hole through the cathode and substrate, exposing the anode.
An array of such wells forms a detector.  The active tracking volume is bounded by a drift
electrode on one side and the wells on the other.  Ionization electrons produced by the passage of
a high-energy charged particle drift toward the anodes and into the wells.  An ionization avalanche occurs
in each well, where there is an intense electric field set up by the voltage applied between the anode 
and cathode.  The electrons from the avalanche are collected on the anode, while an equal but opposite
image charge is measured on the cathode.  

Micro-well detectors have many advantages over conventional multi-wire proportional counters.  The anodes and
cathodes are rigidly affixed to the substrate, which simplifies construction and allows very fine pitch
without electrostatic distortions.  The detectors are very rugged and low cost, and the high intrinsic gain
reduces the power required for readout electronics.  The gain and the drift velocity are independently 
adjustable, allowing high gain to be combined with slow drift; this permits the detectors to be operated
as time projection chambers.  These properties make it practical to read out large gas volumes with very good
three-dimensional position resolution at reasonable cost.

Goddard Space Flight Center has been producing micro-well detectors for the past several years on 
rugged polymide substrates using UV laser ablation\cite{black00,jones1}.  These detectors have been
constructed in a crossed-strip geometry, with the cathodes and anodes forming orthogonal strips on
opposite side of the substrate, to allow simple two-dimensional imaging at X-ray energies.  Very good
detector performance has been observed\cite{black00,jones1,jones2}, including stable operation at gas gains of
$3 \times 10^4$   in Ar- and Xe- based gases, the ability to sustain repeated breakdowns without damage,
a FWHM energy resolution of 18\% for 5.9 keV X-rays in 95\% Xe/5\% CO$_2$ gas, and a 
FWHM position resolution of $\sim 200$ $\mu$m for 400 $\mu$m well pitch.

\begin{figure}
\begin{center}
\begin{tabular}{lr}
\includegraphics[height=7cm]{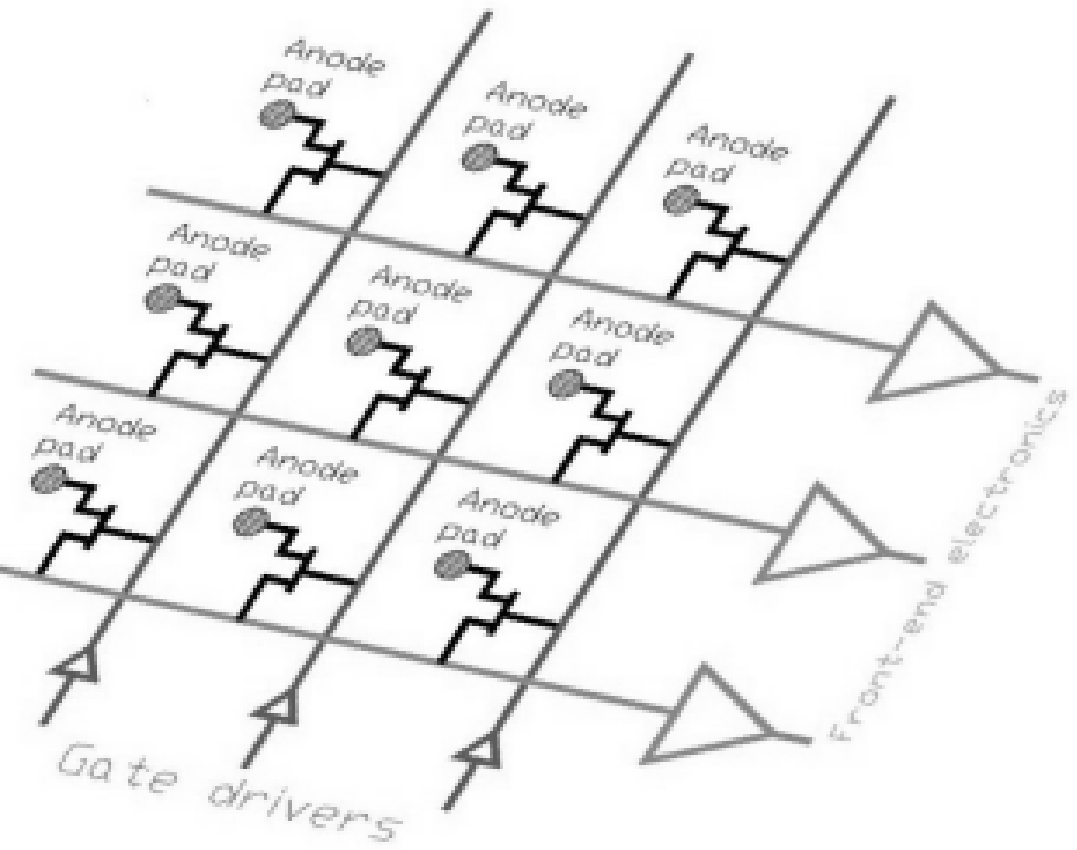}
& \includegraphics[height=7cm]{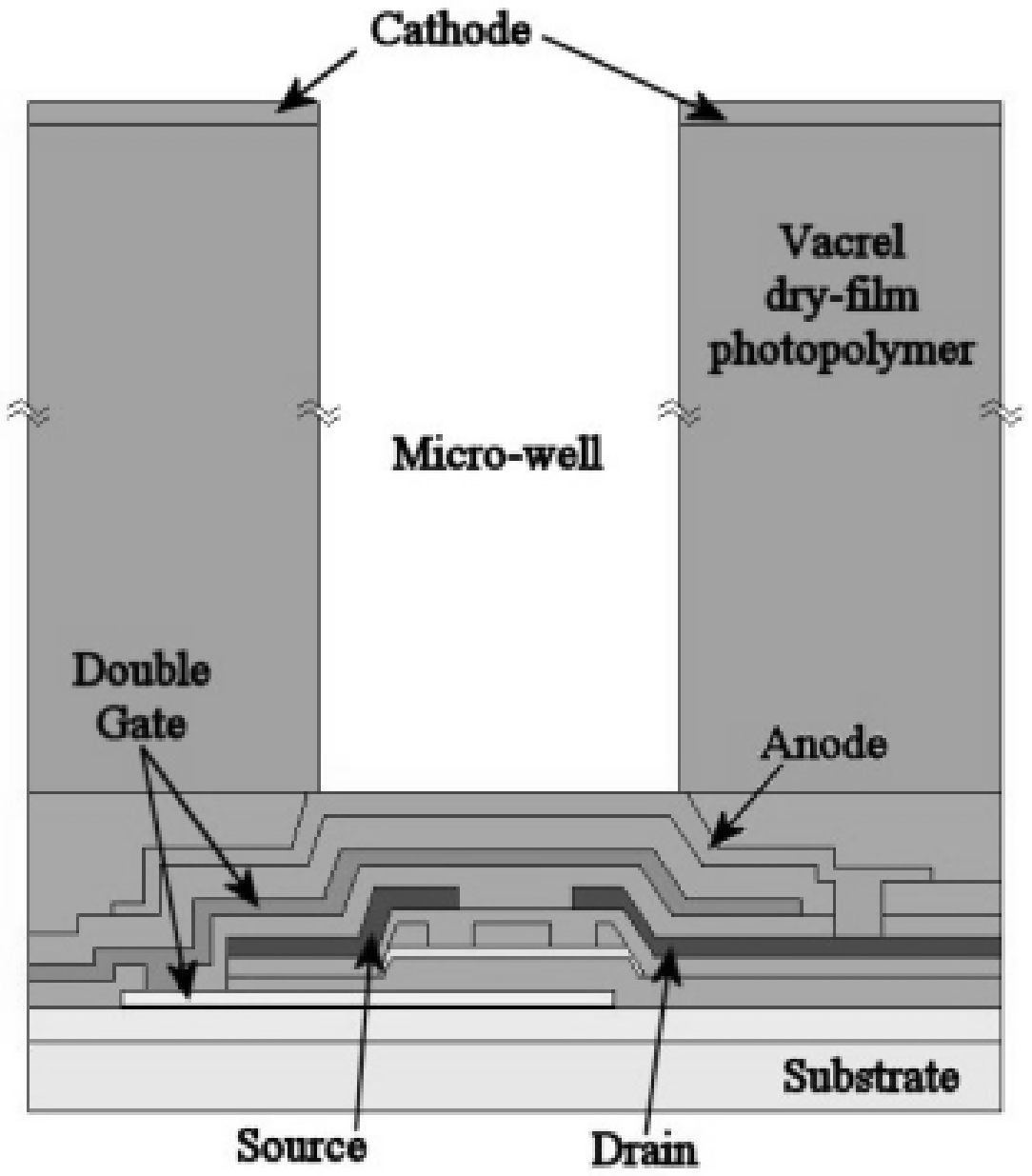}
\end{tabular}
\end{center}
\caption[example] 
{ \label{fig:tft} 
The pixelized micro-well detector concept.  {\em Left:} Readout of a pixelized detector using a TFT
array.  The gate drivers are activated sequentially, allowing each column to be read out in turn by
the charge-integrating front end electronics.  {\em Right:} Cross section of a single integrated
a-Si:H TFT/micro-well detector pixel.  The drift electrode is not shown.  The vertical scale is exaggerated.}
\end{figure}

The crossed-strip readout used for X-ray imaging is not ideal for recording extended tracks from
charged particles.  Thus we are currently focused on the development of pixelized micro-well
detectors (PMWDs) together with collaborators at the Pennsylvania State University (Figure~\ref{fig:tft}).  
The anodes are 
segmented into individual pixel pads, each of which is connected to an element of a thin-film transistor
(TFT) array.  The individual transistor gates are connected in columns, and the outputs are connected in
rows.  The gate drivers for each column are then activated sequentially, allowing the charge collected
on the anode pads to be read out by charge-intergating amplifiers at the end of each row.  Thus a 
two-dimensional projected image of the charged particle track is recorded.  The third dimension may
be determined by measuring the drift time of the ionization electrons (see Section~\ref{sect:apt}).

Our collaborators at Penn State have developed a tri-layer fabrication process for hydrogenated 
amorphous silicon (a-Si:H) TFTs and integrated TFT circuits on polyamide substrates.  TFT arrays based
on a-Si:H were first developed as drivers for flat-panel displays, and more recently as readout devices
for medical imagers\cite{antonuk}.  The current TFT structure consists of a 500 \AA\ thick
amorphous silicon active layer and 3000 \AA\ thick silicon nitride layers as the gate dielectric
and passivation layers.  The structure includes a double gate to reduce leakage current and prevent the
FET from turning on due to the electic field of the micro-well avalanche.  Good electical performance
has been observed, with $I_{on}/I_{off} \sim 10^8$ and a low threshold voltage of 0.8--2.5 V.  Current work
is focused on producing the micro-well detector and TFT arrays as one integrated unit (Figure~\ref{fig:tft}, right) 
using 100 $\mu$m thick-film photolithographic processing.  The cathode and anodes are patterned on the film,
and the wells are created by etching.  Integrated PMWD-TFT detectors with a pitch of 200 $\mu$m 
are currently being produced for testing.

The small pitch and large area of PMWDs imply a very high density for the readout electronics and 
interconnects. The front-end charge-integrating preamplifiers, gate drivers, and time-projection
electronics will all require $\sim 50$ channels cm$^{-1}$ for the detectors described above.
This requires an ASIC implentation for the electronics.  Two charge-integrating amplifiers in an
ASIC package currently under consideration for the PMWDs are the IDEAS VA-SCM2 chip and the Indigo
ISC9717 chip.  We are also investigating tape-carrier packaging and tape automated bonding as means to
achieve our high-density interconnects in a compact, modular manner appropriate for a space-flight
instrument.

\subsection{Advanced Pair Telescope Concept}
\label{sect:apt}

\begin{figure}
\begin{center}
\begin{tabular}{lr}
\includegraphics[height=7cm]{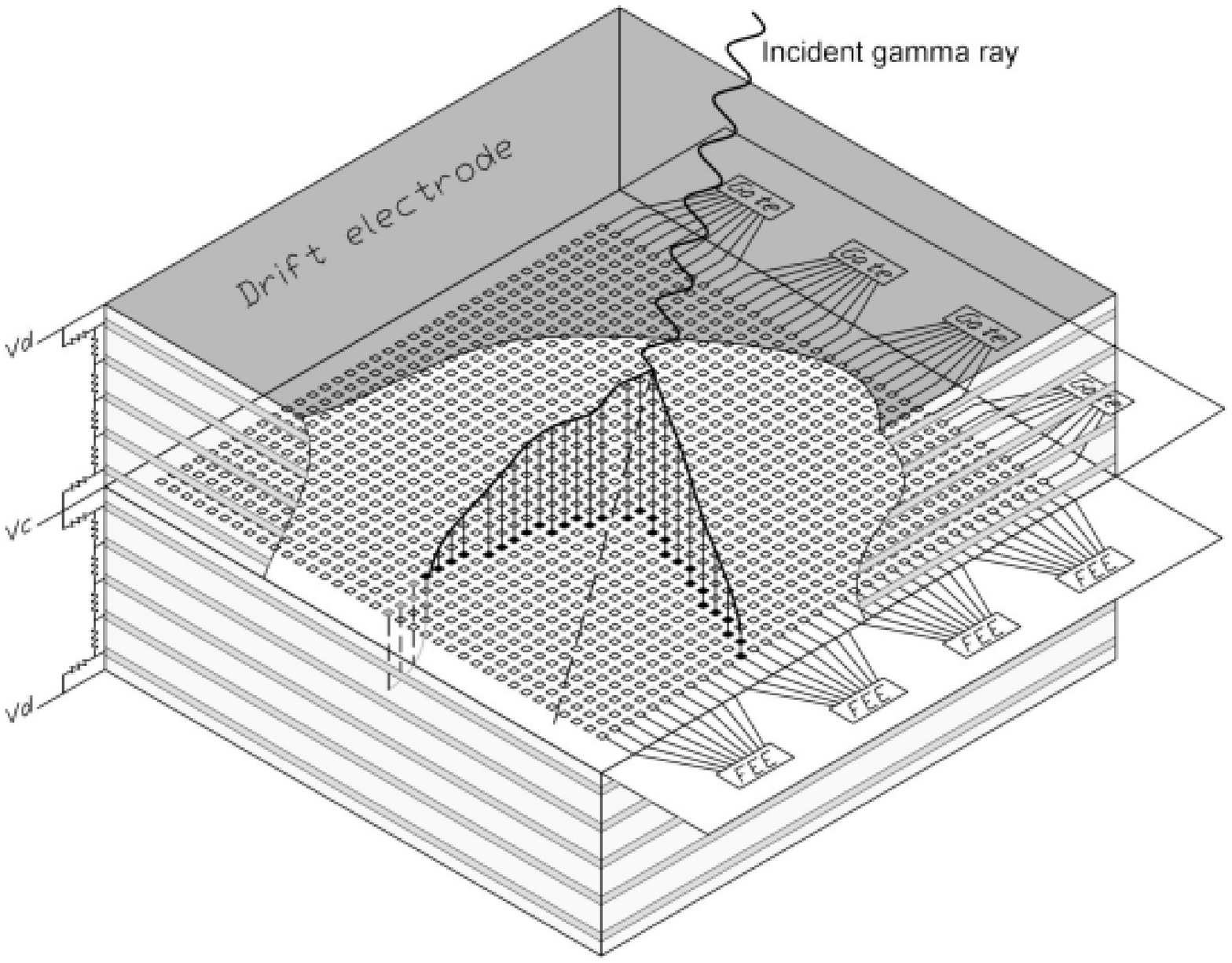}
& \hspace{1.5cm}\includegraphics[height=7cm]{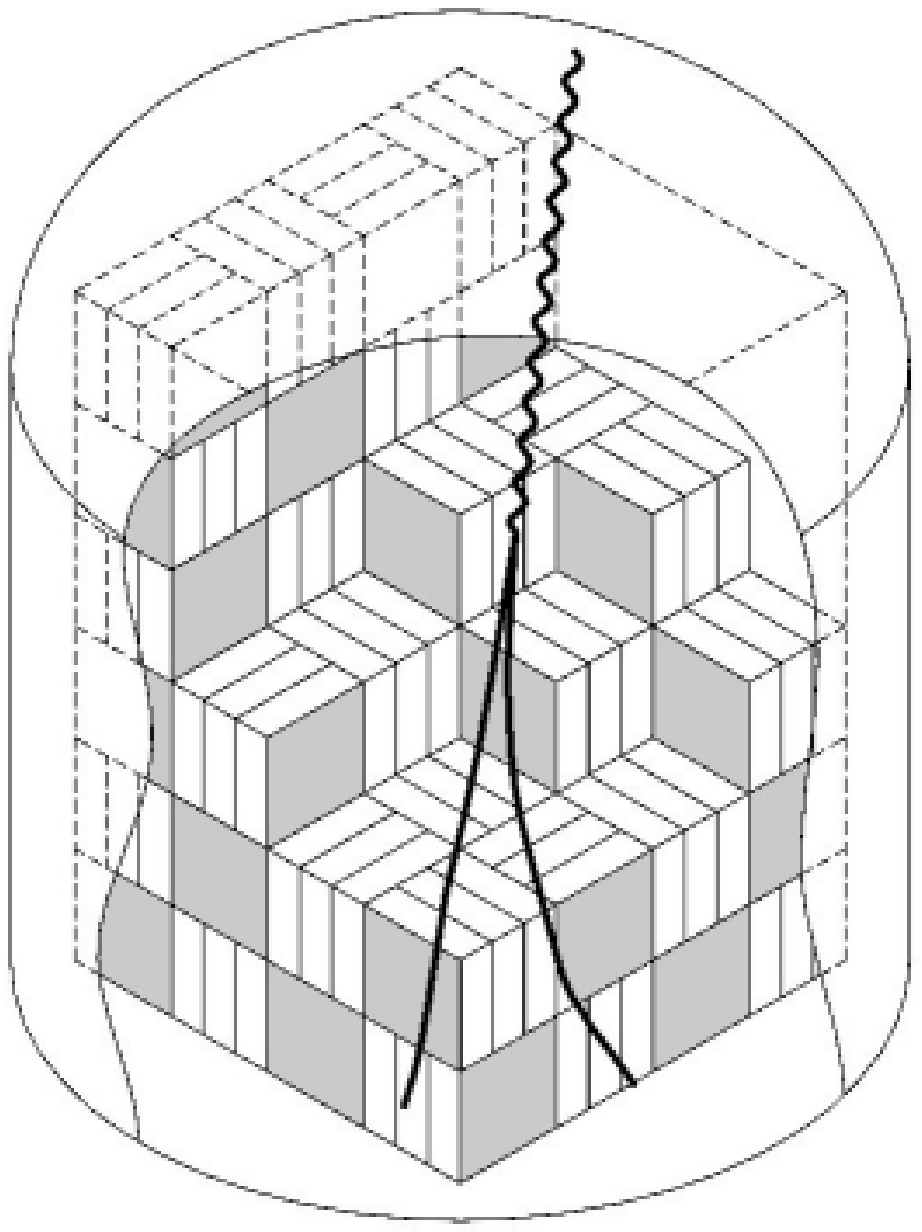}
\end{tabular}
\end{center}
\caption[example] 
{ \label{fig:apt} 
{\em Left:} 
The TTID is based on pixelized micro-well detectors.  Each
individual TTID consists of two back-to-back PMWD-TFT arrays (indicated by the
array of squares), their front end electronics (FEE) and gate driver electronics
(Gate), two drift electrodes (top and bottom), and four support walls.  The walls are ringed with
field shaping electrodes. The drift timing
electronics (not shown) are located on the same circuit board as the gate
driver electronics.
{\em Right:} Instrument concept for the Advanced Pair Telescope.
The APT instrument consists of a number of individual
TTID modules, stacked and arranged in layers, with
the entire stack enclosed in a pressure vessel.  No calorimeter is required since the
electron and positron energies may be measured via their multiple scattering.
}
\end{figure}

The APT imager concept is driven by the requirement to achieve a density of
less than $2 \times 10^{-5}$ RL per sample (Section~\ref{sect:gas}).  
The PMWD-TFT array detectors described above 
with 200 $\mu$m well pitch (and corresponding drift
distance resolution) and a xenon gas mixture at a pressure of $P \sim 1.5$ atmospheres 
(RL $\sim 15 \times (1$ atm$/P$) m), achieves this density 
requirement.  
These PMWD-TFT arrays would be packaged into three-dimensional track imaging detector
(TTID) modules (Figure~\ref{fig:apt}, left).
The APT track imager concept (Figure~\ref{fig:apt}, right)
consists of a large number of TTID modules.  
Each module contains
two back-to-back imagers bounded by drift electrodes and field shaping
electrodes on the four walls.  The front-end, gate driver, and drift
timing electronics are distributed around the periphery of the module and
folded and attached to the walls of the module.  

Photons undergo pair production in the active gas volume of the TTID, defined
by the TFT array and the drift electrode.  The electron and positron traverse
the gas leaving a trail of ionization that drifts into the array of
micro-wells.  The electron and positron also excite the xenon gas, producing UV
scintillation light ($\lambda \sim 180$ nm) that provides a prompt start signal for the drift timing
and, combined with the signals from an anticoincidence detector, the event trigger.  

The low density necessarily requires a large gas depth to obtain a reasonable
pair conversion probability and instrument effective area.  For xenon at 1.5
atm, 0.5 RL (as in EGRET) corresponds to a depth of 5 m.  
We therefore envision a pair telescope consisting of stacked layers of TTID
modules in the form of a cylinder 5 m deep and 2.5 m in diamter.  Modules in successive 
layers are rotated by $90^{\circ}$ so that the drift direction alternates, giving
``stereo'' projected images of the pair tracks.  This give a redundant means of
determining the three-dimensional paths of the tracks.  
Even though the gamma-ray incident
direction and polarization angle are determined within about $6 \times 10^{-3} $ RL $\sim 6$ cm of the
vertex, the electron and positron continue to traverse the remainder of the
gas.  Since the entire depth of gas is instrumented, measuring the Coulomb
scattering of the electron and positron can be used to determine their energy
without the need for a massive calorimeter\cite{hunter}.  It may also be possible to use
an outer layer of gas as an anticoincidence layer, removing the need for
a massive plastic shield dome.

\section{PREDICTED GAMMA-RAY POLARIMETER PERFORMANCE}
\label{sect:pred}

Preliminary estimates of the polarization sensitivity of the APT concept described in
Section~\ref{sect:apt} have been made using Geant4\cite{geant4} Monte
Carlo simulations.  These results should be considered tentative, given the experimental
nature of the Geant4 polarized pair production class.  In addition, the simulations were
performed using Geant4 version 5.0, which has recently been shown to produce
electron-positron multiple scattering results with an rms scattering angle $\sim 20$\% lower than
that predicted by Moli\`ere theory.  We therefore consider our results preliminary.

\subsection{Geant4 Simulations of Polarized Pair Production}
\label{sect:geant}

\begin{figure}
\begin{center}
\begin{tabular}{ll}
\includegraphics[height=7cm,width=8.5cm]{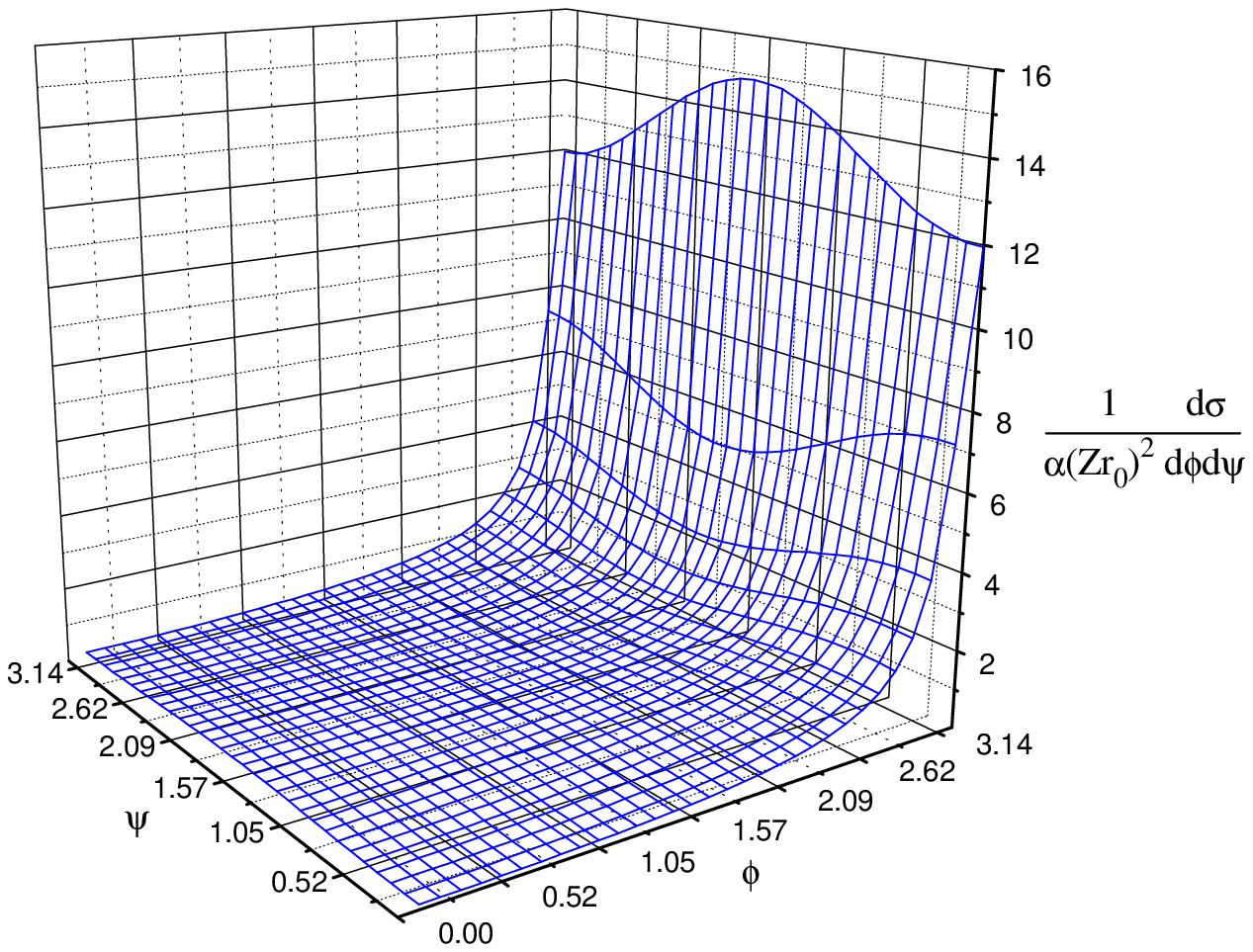}
& \includegraphics[height=7cm,width=7.5cm]{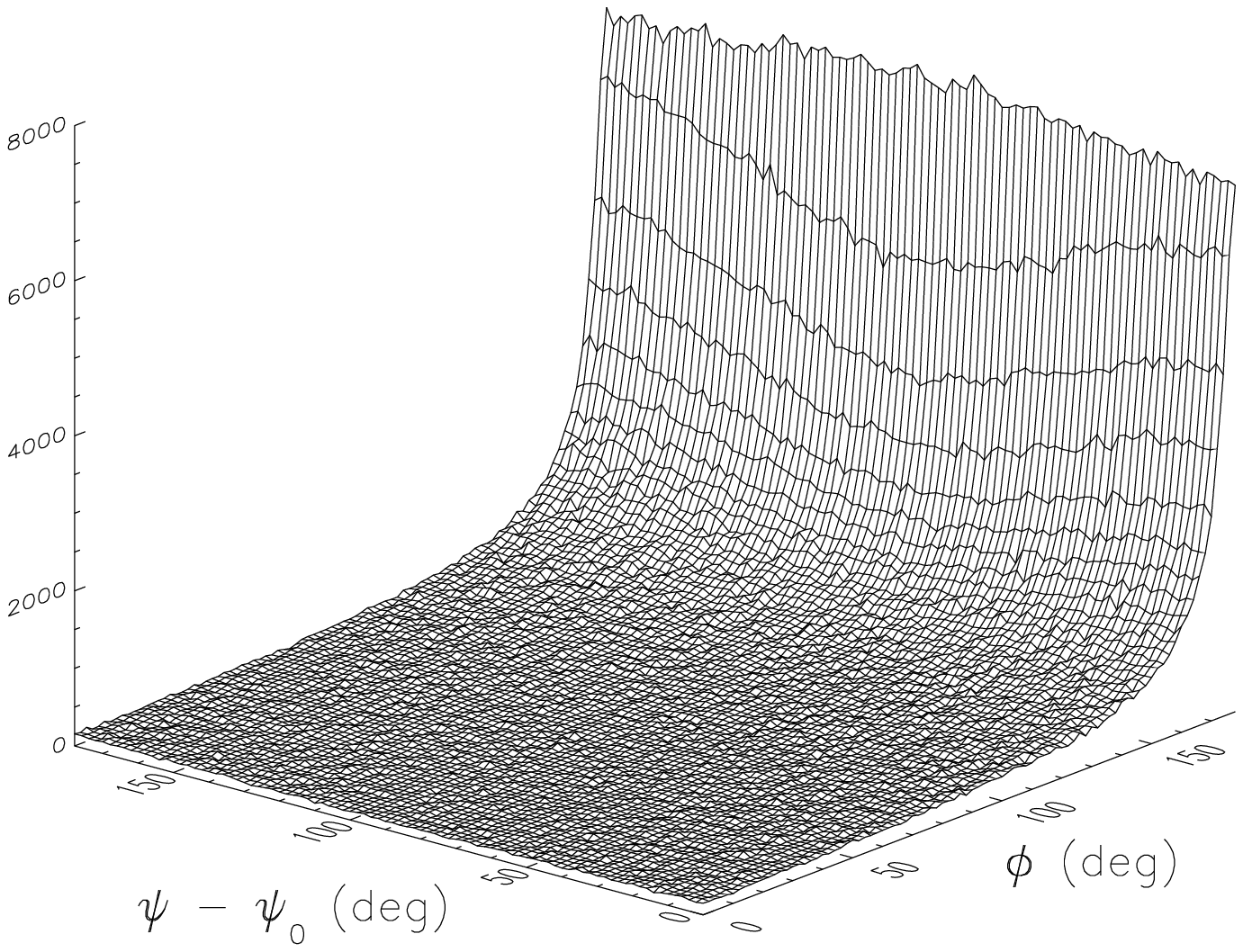}
\end{tabular}
\end{center}
\caption[example] 
{ \label{fig:surf} 
The shape of the polarized pair production cross section surface implemented in Geant4.  
{\em Left:} The cross section surface calculated by Depaola\cite{depaola98}.  Note that
for $\phi \sim 180^{\circ}$ the most likely $\psi$ angle is perpendicular to the polarization vector
(i.e., $\psi - \psi_0 = 90^{\circ}$), whereas for even slightly non-coplanar events the most likely
$\psi$ is parallel to the polarization vector.
{\em Right:} A simulation
was performed using $5 \times 10^6$ events, and a two-dimensional histogram was made of the resulting
$\psi$ and $\phi$ angles with $2^{\circ}$ bins to verify the form of the cross section.  The results are
in good agreement with the calculated cross section.
}
\end{figure}

Polarized pair production has been implemented in Geant4 by G. Depaola and F. Longo\cite{depaola98,depaola99,depaola00}.
The cross section for the pair production process, including the polarization of the incident photon and
the effects of non-coplanarity, has been derived analytically and calculated numerically.  Integrating
over the energies and polar angles of the pair, this yielded the cross section as a function of 
the azimuthal offset from the polarization vector $\psi - \psi_0$ and the non-coplanarity angle $\phi$,
defined as the angle between the electron and positron momenta projected into the plane normal to the
incident photon (i.e., $\phi = 180^{\circ}$ for coplanar events).  This calculated cross section surface is
shown in Figure~\ref{fig:surf} (left).  An analytic parameterization of this surface
is used for the Geant4 Monte Carlo sampling.  Rather than the simple functional form of Equation~\ref{eq:az},
the parameterization takes the form $f(\phi,\psi) = f_{\pi/2}(\phi)\sin^2\psi + f_0(\phi)\cos^2\psi$, where 
$f_{\pi/2}(\phi)$ and $f_0(\phi)$ are functions of $\phi$ whose coefficients are themselves functions
of energy.  The parameterization agrees with the numerically calculated surface to within 5\%.
The cross section surface as implemented within Geant4 is shown in Figure~\ref{fig:surf} (right), 
recovered from a simultaion 
using $5 \times 10^6$ events.
This surface is simply a two-dimensional histogram of the recorded $\psi$ and $\phi$ angles
(with $2^{\circ}$ bins), ignoring the effects of multiple scattering.  
The results of the simulation (Figure~\ref{fig:surf} right) are in good agreement with the calculated 
cross section\cite{depaola99} (Figure~\ref{fig:surf} left),
verifying that the Geant4 polarized pair production class is functioning as expected.
Note that for very nearly coplanar 
events ($\phi \sim 180^{\circ}$) the most likely $\psi$ angle is perpendicular to the polarization vector
(i.e., $\psi - \psi_0 = 90^{\circ}$), but that for even slightly non-coplanar events the most likely
$\psi$ is parallel to the polarization vector, as pointed out by Depaola\cite{depaola99}.  In practice
a telescope would record the maximum at $\psi - \psi_0 = 0^{\circ}$, since multiple scattering and
imperfect position resolution would mask the azimuthal shape at $\phi \sim 180^{\circ}$.

Simulations were performed for the simple case of 100 MeV gamma rays with a fixed
polarization vector entering a xenon gas volume
5 m deep.  No passive material was included in the mass model.  
During the tracking of the charged pair particles the step size was constrained to be 100 $\mu$m, or 
less than the pixel pitch, since otherwise energy was only deposited when a $\delta$-ray with energy
greater than 1 keV was generated.
No detector response
was included in the simulation; rather, the raw output (the exact positions of all energy deposits)
was dumped into an output file, and  
the response of the detectors was applied in a separate program.  This response included 
the effects of electron diffusion in the xenon gas\cite{peisert}, binning into well pixels with a pitch of
200 $\mu$m, and the expected
error in the drift distance, assumed to be equal to the position resolution of
the detectors (taken to be one-half the well pitch, or 100 $\mu$m).  Finally, a simple event reconstruction
algorithm was applied: the two longest, straightest tracks were fit with a straight line for as 
long a distance as
possible, and this was used to determine the infall direction and azimuth angle of the pair plane.  
A simple energy-weighting of the infall direction, according to the inverse of the rms scattering along each
track, has been implemented, but this produced only a small improvement in the angular resolution.

\begin{figure}
\begin{center}
\begin{tabular}{c}
\includegraphics[height=7cm]{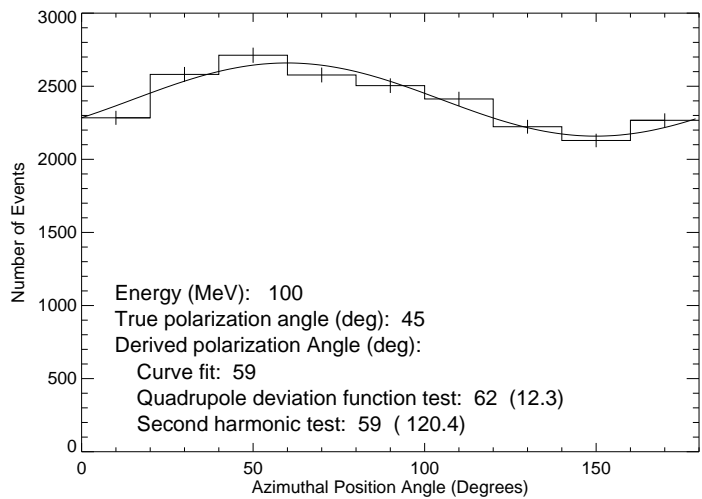}
\end{tabular}
\end{center}
\caption[example] 
{ \label{fig:rawout} 
Output of the Geant4 polarized pair production process for 100 MeV photons in xenon gas.  The effects
of Coulomb scattering and detector binning are not included.  The derived polarization angle found
by curve fitting is 
59$^{\circ}$, whereas the input angle was 45$^{\circ}$; the difference is due to the non-coplanar
events.  The modulation factor is $Q_{100} = 0.104$.
The quadrupole deviation function and second harmonic tests also find an angle of $\sim 60^{\circ}$; 
the values of the test statistics are 
given in parentheses.
}
\end{figure}

Figure~\ref{fig:rawout} shows the output of the Geant4 pair polarization process.  
Events were
selected from the raw event file, taking the exact positions of all energy deposits and using only the
first two hits of the electron and positron tracks (100 $\mu$m from the vertex) to define the initial
pair momenta.  Thus the effects of Coulomb scattering and detector binning were removed and only the effect
of the parameterized cross section should be present.  Only events falling within an angular radius
containing 68\% of all events were included.  The input polarization angle was 45$^{\circ}$.

The presence of an azimuthal modulation was searched for in several ways.  
The first is a simple function fitting
as described in Section~\ref{sect:principles}.  This yielded a polarization angle of $\psi_0 = 59^{\circ}$ 
and a modulation factor $Q_{100} = 0.104$.  
The second method is the ``quadrupole deviation function test'' used to search COS B data for
evidence of polarization from the Vela Pulsar\cite{caraveo,mattox90}.  The test statistic must exceed
4.1 for a $3\sigma$ detection of azimuthal asymmetry.  The test statistic for
this dataset, given in parentheses in Figure~\ref{fig:rawout}, is 12.3, a very significant detection.
The polarization angle is found to be $62^{\circ}$ by this method.
The third method is the even more sensitive ``second harmonic test'' used to
search for polarization signals in COS B and EGRET data\cite{mattox90,mattox91}.  The value of the
test statistic must exceed 13.8 for a $3\sigma$ detection.  The test statistic for
this dataset is 120.4, also a very significant detection.  The
polarization angle is found to be $59^{\circ}$ for this test as well.  

The shift in the polarization angle
found by all three methods is the result of the non-coplanar events.  The angle $\psi$ used in the
cross section is actually the azimuth angle of the momentum of the 
pair electron, projected into the plane perpendicular to the initial photon direction\cite{depaola99}, 
whereas a telescope can only measure the angle formed by the plane of the electron and positron.  For 
$\phi \sim 180^{\circ}$ these angles are nearly the same, and this is the case for which Equation~\ref{eq:az} 
applies.  For $\phi \lesssim 170^{\circ}$, however, the angles are significantly different, and since
a real telescope cannot recognize and exclude non-coplanar events, these events will systematically
shift the azimuthal distribution recorded.  In practice, since the shape of the cross section surface
has been calculated as a function of energy, it should be possible to correct for this effect.

\subsection{Expected Polarization Sensitivity}
\label{sect:sens}

\begin{figure}
\begin{center}
\begin{tabular}{c}
\includegraphics[height=7cm]{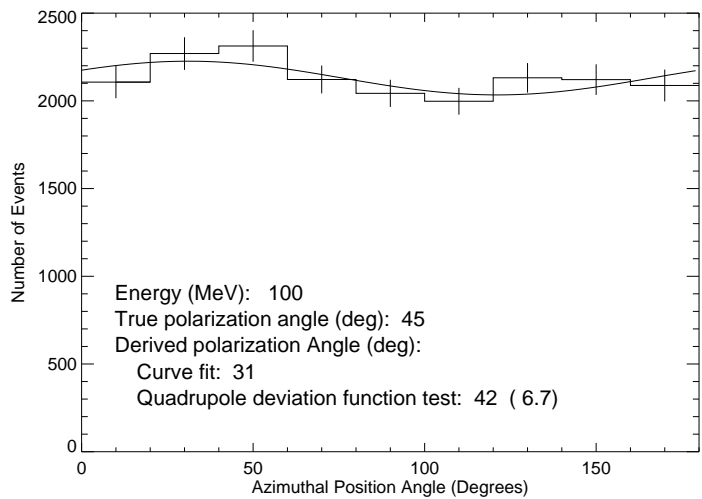}
\end{tabular}
\end{center}
\caption[example] 
{ \label{fig:subpol} 
Azimuthal distribution obtained from Geant4 after applying detector response and event
reconstruction routines.  Only event within the 68\% angular radius are included, given
a total of $\sim 1.9 \times 10^4$ events.  The initial polarized distribution was divided
by a normalized unpolarized distribution to remove systematics effects.  The modulation
factor is $Q_{100} = 0.045$.
}
\end{figure}

Out of $1.35 \times 10^5$ input events at 100 MeV, $3.0 \times 10^4$ produced pair events that
were successfully reduced
by the simple reconstruction algorithm, given an effective efficiency of $\sim 22$\%.  
(The reconstruction algorithm had trouble fitting events that included extended delta rays or
very uneven energy splits between the electron and positron.)
Only events within the
68\% angular radius were included in the azimuthal histogram, as this would be the case for a 
real observation.  This gave a total
of $\sim 1.9 \times 10^4$ events.  
The 68\% angular radius found using energy weighting of the pairs was $1.3^{\circ}$.  
In order to recover the azimuthal distribution
of the pairs it was necessary to repeat the simulation and reduction using unpolarized photons, and
then to divide the polarized azimuth histogram by the normalized unpolarized histogram.
The azimuthal distribution thus obtained
is shown in Figure~\ref{fig:subpol}.  
Although the statistics are poor, a modulation similar to that in Figure~\ref{fig:rawout} is 
evident.  Due to the binning required to perform the division by the unpolarized histogram,
it was not possible to apply the second harmonic test to this data (the second harmonic test 
operates on the individual azimuth values to increase sensitivity\cite{mattox90}).  Both the
curve fit and quadrupole deviation function tests find significant azimuthal asymmetry (the 
test statistic for the quadrupole deviation function test is 6.7), although
they are unable to recover the input polarization angle from Figure~\ref{fig:rawout}.  
Better statistics should improve the determination of the polarization angle (this initial
simulation was limited by time constraints).
The modulation
factor obtained from the curve fit is $Q_{100} = 0.045$.  Thus $R^{\prime}/R \sim 0.45$, implying that
the pair plane is indeed being measured within the $6 \times 10^{-3}$ RL required from 
Section~\ref{sect:gas}.

Using the results of the simulation, we may estimate the $3\sigma$ MDP from Equation~\ref{eq:mdp}.
We assume a diameter of 2.5 m and a detection efficiency of 22\%, giving an effective area of
10800 cm$^2$.  Only events within a 68\% angular radius of $1.3^{\circ}$ are included.  The background
is taken to be events from the observed EGRET extragalactic diffuse background spectrum\cite{kniffen} 
which fall within the 68\% angular radius.  We consider a Crab-like source, and, for simplicity,
assume a 100 MeV-wide energy band centered on 100 MeV.  Then the $3\sigma$ MDP for a 1 Crab
source is 40\% in 10$^6$ s of observation time.  In 10$^7$ s of observing time, or about one year
of real time, the MDP is 13\%.  The MDP for a 100 mCrab source in 10$^7$ s is 41\%.  These numbers,
while preliminary, indicate that this APT concept will have a useful polarization sensitivity for
bright sources around 100 MeV.

\section{Conclusions}

Our preliminary Geant4 simulations of polarized pair production indicate that an APT consisting of
PMWD-TFT gas detectors will be capable of detecting linearly polarized emission from bright sources
at 100 MeV.  There are several possible methods of improving the sensitivites derived in 
Section~\ref{sect:sens} that we will explore in more detailed simulations.  These include limiting
the effects of drift electron diffusion by shortening the drift distance, and improving the
event reconstruction algorithms.  Due to the inherent weakness of the azimuthal asymmetry for
polarized pair production, however, the most effective means may be simply to increase the number
of events available.  Since the angular resolution of a gas pair telescope is exceptionally good,
including events out to greater than the 68\% angular radius used for EGRET should not increase
the background too much.  It may also be possible to increase the diameter of the telescope, 
especially if a plastic anticoincidence shield is not required.  For example, a 3 m diameter and
inclusion of 90\% of the events would improve the $3\sigma$ MDP for a 1 Crab source to 29\% in
10$^6$ s.

Another possible improvement is the inclusion of triplet pair production events.  Pair production
can occur in the field of an electron in the conversion material as well as in the field of the
nucleus, and the recoil momentum imparted to the electron will be modulated around the polarization
vector in the same manner as the pair plane\cite{boldyshev}.  This recoil electron may be tracked
in a gas detector, and the azimuthal distribution will be far easier to determine than that of the
pair planes.  The cross section for this process is reduced by a factor of $1/Z$ compared to 
nuclear pair production, so that it only occurs $\sim 2$\% of the time in xenon.  Sensitivity to this
process may therefore require lighter gases.  Triplet production is not yet implemented in Geant4.

We will continue to improve our Geant4 mass model, detector response, and event reconstruction
algorithms in order to explore the polarization sensitivity of our APT concept more thoroughly.  
We also plan to construct a prototype, together with our Penn State collaborators, and to test it
at a polarized gamma-ray beam in 2006.

\acknowledgments     
 
This work was performed while the author held a National Research Council Research
Associateship Award at NASA/GSFC.


\bibliography{spie2003_paper}   
\bibliographystyle{spiebib}   

\end{document}